\documentclass[12pt]{article}
\usepackage{graphicx}
\usepackage{amssymb}
\begin{document}

\title{On a Modified Klein Gordon Equation}
\author{ B S Lakshmi\footnote{email:bslakshmi2000@yahoo.com}\\ Department of Mathematics\\JNT University, Kukatpally \\Hyderabad}
\date{}
\maketitle
\begin{abstract}
We consider a modified Klein-Gordon equation that arises at ultra high energies. In a suitable approximation it is shown that
for the linear potential which is of interest in quark interactions, their confinement for example,
 we get solutions that mimic the Harmonic oscillator energy levels, surprisingly. An equation similar to the beam equation is
 obtained in the process.
\end{abstract}

% ----------------------------------------------------------------
\section{Introduction}
If we start with the Snyder-Sidharth Hamiltonion
\cite{gl-ap09}, we get a fourth order differential equation
which is very complicated but interesting- as we will see, such an
equation also features in the theory of beams and robotic arms. The
Hamiltonian is
\begin{equation} \label{sny-sid-ham}
E^2 = m^2 +p^2 + \alpha l^2 p^4 \end{equation} In equation
(\ref{sny-sid-ham}) $E$ is the relativistic energy, $m$ the rest
mass of the particle,$p$ its momentum, $\alpha$ a dimension-less
constant of the $0(1)$ and $l$ is a fundamental minimum length
typically a Compton length (including as a special
case the Planck length) \cite{bg-uof05} and we work
with natural units in which $c=1=\hbar $. Equation
(\ref{sny-sid-ham}) leads to a modified Klein-Gordon equation,
\begin{equation}\label{modkg}
(\alpha l^2 \nabla ^4 +\square - m^2) \psi=0
 \end{equation}
 In Equation (\ref{modkg}) the $\square$ denotes the D'Alembertian,
 given by
$$
\square= \frac{\partial^2}{\partial t^2}-\frac{\partial^2}{\partial
x^2}-\frac{\partial^2}{\partial y^2}-\frac{\partial^2}{\partial z^2}
$$

 In case there is a potential $U$, this has to be inserted in the parenthesis of Equation (\ref{modkg}).
 Equation (\ref{modkg}) is a formidable equation, but we will see below that with simplifications, we can get some interesting results.
 \section{Approximate Solutions}
 As in the usual theory, we consider solutions which can be separated into the space and time parts \cite{pocr-61}.
  We further consider for simplicity, the case of one space dimension, because it already gives interesting conclusions.
  Thus our starting point is
 \begin{equation}\label{sneq-3}
 \alpha l^2 \frac{d^4 \psi}{dx^4} - \frac{d^2 \psi}{d x^2}- (E^2+m^2+U(x))\psi = 0
 \end{equation}
 We consider different cases for the potential $U(x)$.

 \noindent
 \\ Case (1). \\
 $U(x) =0$
 Then we canwork out that the solution of Equation (\ref{sneq-3}) is given by
   \begin{large}
  \begin{eqnarray*}\psi(x) &=& e^{\frac{x \sqrt{\frac{1}{l^2 \alpha }-\frac{\sqrt{4 \left( E ^2+m^2\right) \alpha  l^2+1}}{l^2 \alpha }}}{\sqrt{2}}}
   c_1+e^{-\frac{x \sqrt{\frac{1}{l^2 \alpha }-\frac{\sqrt{4 \left(E^2+m^2\right) \alpha  l^2+1}}{l^2 \alpha
   }}}{\sqrt{2}}} c_2\\ & &+ e^{x \sqrt{\frac{\sqrt{4 \left(E^2+m^2\right) \alpha  l^2+1}}{2 l^2 \alpha }+\frac{1}{2 l^2 \alpha
   }}}c_3+e^{-x \sqrt{\frac{\sqrt{4 \left(E^2+m^2\right) \alpha  l^2+1}}{2 l^2 \alpha }+\frac{1}{2 l^2 \alpha }}} c_4
  \end{eqnarray*}\end{large}

  Depaending on the signs of $E^2+m^2$, these could represent bound (or antibound)
  or scattered solutions.\\
   \noindent Case (2).\\ $U(x) =x.$ (or a multiple of $x$.) \\ In this case we can get meaningful results by considering a two-step approximation. First we
  observe that as the term $\alpha l^2$ in Equations (\ref{sny-sid-ham}),(\ref{modkg}) or (\ref{sneq-3}) is very small
  compared to the other coefficients, this being of the order of the square of the Compton length, the fourth derivative term
  can be neglected in the first instance. So the Equation (\ref{sneq-3}) now reduces to a second order equation
  \begin{equation}\label{sneq-4}
 \psi''(x)+ \left(E^2+m^2+x\right) \psi(x)=0
  \end{equation}
  We can easily show that the solutions of Equation (\ref{sneq-4}) are given in terms of the Airy Functions.
  \begin{equation}\label{sneq-5}
  \psi_0(x)=\mbox{AiryAi}\left(\sqrt[3]{-1} \left(E^2+m^2+x\right)\right) d_1+\mbox{AiryBi}\left(\sqrt[3]{-1} \left(E^2+m^2+x\right)\right) d_2
  \end{equation}
  In Equation (\ref{sneq-5}) we consider only the decreasing exponential for meaningful solutions.
From the well-known asymptotic approximation \cite{lasa-70},
 of the AiryAi function the decreasing exponent part of this
 solution can be shown to be
 $$\frac{1}{2 \sqrt \pi} e ^{\frac{i \pi}{12}}(-1)(E^2+m^2+x)^{-1/4}e^{\frac{2}{3}i (E^2+m^2+x)^{3/2}}$$
 which is approximately equal to $$(E^2+m^2+x)^{-1/4}e^{\frac{2}{3}(E^2+m^2+x)^{3/2}}$$
 Now inserting the solution of Equation (\ref{sneq-5}) in the fourth derivative of Equation (\ref{sneq-3}),that is differentiating the solution obtained in Equation (\ref{sneq-5}) four times  we get
 \begin{equation}\label{sneq-6}g(x)= \frac{e^{-\frac{2 x^{3/2}}{3}} \left(32 \left(8 x^6-16 x^{9/2}+x^3+10 x^{3/2}\right)+585\right)}{256 x^{17/4}}\end{equation}
  which can be treated as an inhomogeneity term to  be added to Equation (\ref{sneq-4}),

 While this gives a better approximation at the next level, it is still too complicated  for computational purposes. To simplify matters further, we observe that for large $x$, the term $x^2 \psi _0$ in Equation (\ref{sneq-6}) dominates. Inserting this into Equation (\ref{sneq-4}) we get, \begin{equation}\label{sneq-7}
\psi_{1}''(x)+ \left(E^2+m^2-c x^2+x\right) \psi_{1}(x)=0,
 \end{equation}
 where $c$ is a constant.
 The solution of Equation (\ref{sneq-7}) can be shown to be
 \begin{equation}\begin{array}{llll}\label{sneq-8}
 \psi_{1}(x)&=&e^{\frac{x-c x^2}{2 \sqrt{c}}} c'_1\, \mbox{H}\left[{\frac{4 c E^2+4 c m^2-4 c^{3/2}+1}{8 c^{3/2}}},
 \sqrt[4]{c} x-\frac{1}{2 c^{3/4}}\right]+
 \\& &
 e^{\frac{x-c
   x^2}{2 \sqrt{c}}} c'_2 \, \mbox{F}_1\left[-\frac{4 c E^2+4 c m^2-4 c^{3/2}+1}{16 c^{3/2}},\frac{1}{2},
   \left(\sqrt[4]{c} x-\frac{1}{2
   c^{3/4}}\right)^2\right]
 \end{array}\end{equation}
 where H stands for the Hermite polynomial and F$_1$ the Hypergeometric function of the first kind.
 In Equation (\ref{sneq-8}) we consider, for meaningful solutions the part that consists of the Hermite polynomial and the
 decreasing exponent.
 The energy levels follow by applying the condition that these solutions should also vanish
 at the origin.

  It will be immediately seen that this solution resembles the usual Harmonic oscillator solutions.
 Indeed this should not be surprising, because for large $x$ in Equation (\ref{sneq-7}) the term $x^2$ dominates. \\
 It is interesting to note that if in Equation (\ref{sneq-3}), we substitute in the term $\frac{d^2\psi}{dx^2}$, the
 approximate solution $\psi_0$ or $\psi_1$, then we get the equation,
 \begin{equation}\label{sneq-9}
 \frac{d^4 \psi}{dx^4} -  \psi=h(x).
 \end{equation}
Interestingly, Equation (\ref{sneq-9}) occurs in the theory of beams
\cite{weli-99}. The Airy function itself has been considered by the
author elsewhere \cite{lara-07}.
 \section{Conclusion}
What we have shown here is that for the modified Klein-Gordon
equation (\ref{modkg}), which follows from the Snyder-Sidharth
Hamiltonian, meaningful solutions exist for the free particle as
also for a particle in a potential linearly varying as the distance.
 In the latter case, surprisingly the equation approximately mimics the case of the Harmonic oscillator in which the
 potential varies as the square of the distance. All this is important in the context of quark confinement.
  We also get in the process an equation mimicking the behavior of a beam.

\end{document}